\documentclass[aps,prl,twocolumn,superscriptaddress]{revtex4}

\usepackage{graphicx}
\usepackage{morefloats}
\usepackage{color}
\usepackage{amssymb}
\usepackage{float}
\usepackage{bm}
\usepackage{amsmath}
\usepackage{epstopdf}

\begin{document}

\newcommand{\ham}{\mathcal{H}} 
\newcommand{\Q}{\mathcal{Q}} 

\title{Interface moment dynamics and its contribution to spin-transfer torque switching process in magnetic tunnel junctions }

\author{Christopher Safranski}
\author{Jonathan Z. Sun}

\affiliation{IBM T. J. Watson Research Center, Yorktown Heights, New York 10598, USA}


\begin{abstract}
A practical problem for memory applications involving perpendicularly magnetized magnetic tunnel junctions  is the reliability of  switching characteristics at high-bias voltage. Often it has been observed that at high-bias, additional error processes are present that cause a decrease in switching probability upon further increase of bias voltage. We identify the main cause of such error-rise process through examination of  switching statistics as a function of bias voltage and applied field, and  the junction switching dynamics in real time. These experiments show a coincidental onset of error-rise and the presence of a new low-frequency microwave emission well below that dictated by the anisotropy field.  We show that in a few-macrospin coupled numerical model, this is consistent with an interface region with concentrated perpendicular anisotropy, and where the magnetic moment has limited exchange coupling to the rest of the layers. These results point to the important role high-frequency interface magnetic moment dynamics play in determining the switching characteristics of these tunnel junction devices.

\end{abstract}

\flushbottom
\maketitle

\section{Introduction}

Perpendicularly magnetized magnetic tunnel junctions (p-MTJ) are  enabling devices for spin-transfer torque switched\cite{Slonczewski1996,Berger1996,Sun1999} magnetic random access memory (STT-MRAM)\cite{Worledge2011,Ikeda2010,Apalkov2016,2017019}, allowing for more efficient memory-logic integration, and for advanced neuromorphic and other information processing applications beyond the von Neumann architecture\cite{Torrejon2017,Camsari2018}. 

For these applications, an important device characteristics is its switching error probability vs drive-current (voltage) amplitude. This so-called ``write-error-rate" (WER), defined as a write-error probability per switching operation $\epsilon_r$, is a function of the write speed and write voltage across an p-MTJ. This WER behavior originates from STT-dynamics with thermal-agitation. It is important to achieve low WER well below $10^{-6}$/write-operation, at given write speed and voltage, and consistently across all devices in a memory chip\cite{Nowak2011,2016085}. However, due to the symmetric nature of spin-transfer torque versus transport current direction\cite{2005129,2005073,Kim2008}, the same force that switches the so-called magnetic free-layer (FL) in an p-MTJ will also impose a destabilizing force on the reference-layer (RL) following the FL switching. Such behavior causes a second threshold in the write-voltage, above which additional write-error would occur.\cite{Kim2008,Sun2009,Choi2016,Yoshida2018,2019009} To mitigate these issues, it is essential to fully understand the STT-related magnetic dynamics on both sides of the tunnel barrier in an p-MTJ.

\begin{figure}[pt]
\includegraphics[width=0.5\textwidth]{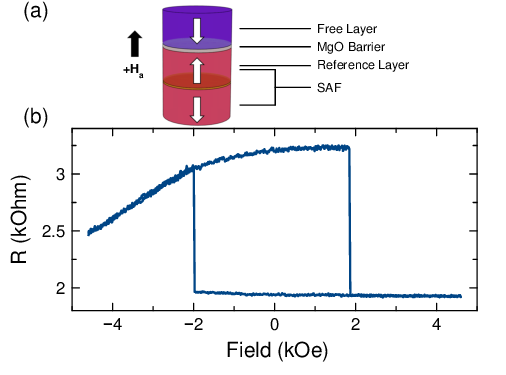}
\caption{(a) Schematic of the magnetic tunnel junction layers. The reference layer is composed of an MgO-CoFeB interface, followed by a synthetic antiferromagnet (SAF) structure.    (b) p-MTJ resistance $R$ as a function of applied magnetic field.}
\label{fig1}
\end{figure} 

\begin{figure*}[tbp]
\includegraphics[width=\textwidth]{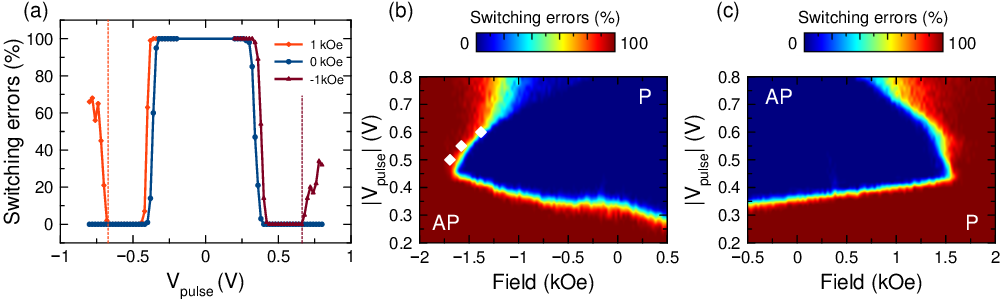}
\caption{ 
(a) WER (in \% per switching operation) {\it vs} pulse voltage height $V_\mathrm{pulse}$ at three different applied magnetic fields. $+V_\mathrm{pulse}$ and $+\mathbf{H}_a$ directions correspond to switching intended for AP-to-P . (b) and (c): Contour plot showing the switching errors for various applied field and pulse magnitudes for AP-to-P (b) and P-to-AP (c) transitions. All measurement are performed with a 100\,ns pulse width. White data points on (b) correspond to the onset field of the LF mode (white dotted line) -- see discussion surrounding Fig.\,\ref{fig_emission}(c).
}
\label{fig2}
\end{figure*} 

STT-induced magnetic switching involves an anti-damping process, with the transport spin decoherence induced spin-torque heavily concentrated near the interface receiving the spin-current.\cite{Slonczewski1996,2005129} This makes it particularly important to understand the role of the interface moments. For CoFeB-MgO-CoFeB type of p-MTJs, it is known both from theoretical\cite{Cuadrado2018,Hallal2013} and from experimental studies\cite{Barsukov2015,Fu2016} that the formation of Fe-O bond near the MgO interface could lead to a difference in properties between the interface moments and the rest of the CoFeB electrodes. Further, the MgO-barrier initiated (bcc) crystallization of FeCo at the tunnel barrier interface during post-deposition annealing\cite{2007079} weakens these interface moment's magnetic exchange across MgO-dictated lateral grain-boundaries. The MgO-FeCo interface is also responsible for concentrated perpendicular magnetic anisotropy (PMA) potential for these interface moments. The presence of MgO tunnel barrier interface concentrated PMA reduces interface moment's rotation in hard-axis field as seen by tunnel magnetoresistance. It also causes an apparent 4th order magnetic anisotropy.\cite{2015095,2019010} It has also recently been related to the observed size and RA dependence of STT-switching threshold in p-MTJ that are clearly non-macrospin in nature.\cite{Sun2017}

In this work, we investigate the switching statistics and real-time dynamics of an p-MTJ as they depend on write-voltage and external bias magnetic field. By examining both in time- and frequency-domain, we experimentally probe the origin of high-bias WER anomaly. It is shown to correspond to a new form of microwave emission with a lower frequency dispersion with magnetic field compared to normal ferromagnetic resonance. We show this is consistent with an STT-driven interface magnetic moment instability  that reduces much of the PMA in the p-MTJ layers adjacent to MgO tunnel barrier. Accompanying FL and RL instabilities then lead to a WER deterioration.

\section{Experimental Results}

In the following sections, our  measurements show that (1) there is a bias-field dependent WER rise at high bias voltage beyond the main switching threshold across the p-MTJ;(2) the onset of such WER rise is systematically dependent on an easy-axis applied magnetic field; (3) This type of WER rise apparently does {\it not} involve a permanent reversal of the entire RL;  and (4) the onset boundary of such WER rise in  applied field and voltage   parameter space coincides with the observation of a new low-frequency microwave emission mode, whose frequency-magnetic field dispersion   shows intercepts much reduced from the FL anisotropy field $H_\mathrm{k}$.

We measure a circular 75\,nm p-MTJ represented in Fig.\,\ref{fig1}(a), patterned from magnetic films consisting of a CoFeB-like free and reference layers separated by an MgO tunnel barrier similar to those discussed in Refs.\cite{Sun2013,Safranski2016}. Figure\,\ref{fig1}(b) shows the p-MTJ's resistance (R) as a function of an applied field  $\mathbf{H}_a$ perpendicular to the sample plane. The p-MTJ free layer can be switched with the application of roughly 2\,kOe of magnetic field. The high field decrease in resistance is related to un-optimized, insufficient PMA at the RL-MgO interface, and may involve partial rotation of the top RL moments in elevated bias fields.

\begin{figure*}[pt]
\includegraphics[width=\textwidth]{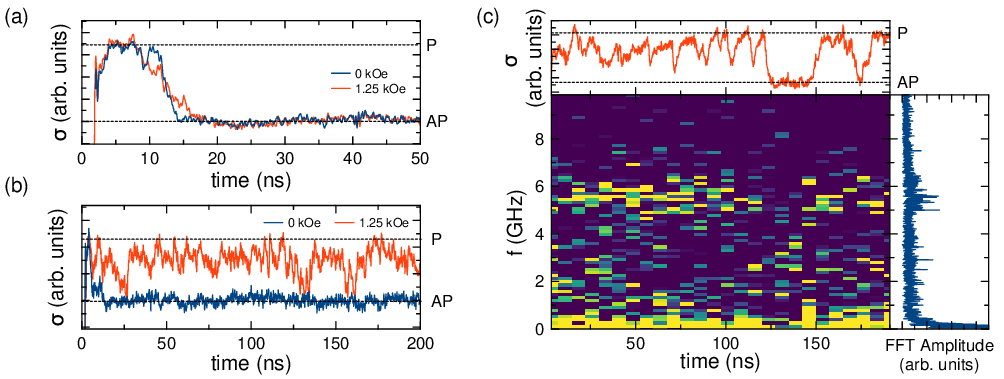}
\caption{
(a) Single shot time traces of the junction conductance when a -0.5\,V pulse is applied at 0 and 1.25\,kOe.   For clarity, traces shown in the figures undergo a moving average of five data points. Dotted lines represent junction conductance in the P and AP states. (b) The junction conductance when a -0.65\,V pulse is applied at 0 and 1.25\,kOe. (c) Time trace (top) of the junction conductance when  0.55\,V pulse is applied along   with a -1.75\,kOe field. The contour plot shows the FFT of the time trace as a function of time with the average plotted on the right. 
}
\label{fig3}
\end{figure*}

\subsection{STT Switching and WER statistics}

We first characterize the device performance with spin transfer torque switching statistics. The p-MTJ is initially reset into a known state, then  a 100\,ns long voltage pulse is applied. The resulting resistance is read to determine if a switch was made. This is repeated 200 times at each fixed pulse amplitude to establish the switching probability, to an accuracy of 0.5\%, of the device at the given pulse width and amplitude. This process is then repeated for various voltage amplitudes and bias field values to map out the WER behavior of a device, as shown in Fig.\,\ref{fig2}.  

Figure\,\ref{fig2}(a) shows that, at zero external bias field, the device switched as intended from anti-parallel to parallel (AP-to-P) around +0.4\,V, and P-to-AP, -0.4\,V, with no visible error upon further increase of $\left| V_\mathrm{pulse}\right|$. However, with -1\,kOe bias field, it shows a WER-rise at $V_\mathrm{pulse}\sim$+0.7\,V and beyond (dashed line), corresponding to error events reverting the p-MTJ into AP-state in bias-direction for switching to P. Similarly, at +1 kOe bias, a WER-rise is seen for the P-to-AP switching beyond $\sim$ -0.7\,V (dashed line). 

This dependence of WER-rise on magnetic field bias is systematically mapped out for both AP-to-P and P-to-AP direction of $V_\mathrm{pulse}$ in Fig.\,\ref{fig2}(b) and (c).   For the AP-to-P transition shown in Fig.\,\ref{fig2}(b),  the initial switching voltage decreases  as the magnetic field is swept towards the positive direction. This is a result of the magnetic field being applied in a direction that encourages switching and thus decreasing the required spin current to induce switching. In contrast, as the magnetic field is swept towards  positive values, the onset  of the high voltage errors increases. For the P-to-AP transition shown in Fig.\,\ref{fig2}(c), a similar trend is seen. The application of negative magnetic fields  to encourage  switching lowers the initial switching threshold  and increases the onset  for high voltage errors.   

We observe that the SAF layer's resting orientation remains unchanged when these erroneous switching events occur  by checking the R($\mathbf{H}_a$) loop direction upon such switching error detection.

\subsection{Time Domain Dynamics}

To elucidate the cause for the high voltage switching errors, we investigate the p-MTJ dynamics in the time domain. A single-shot 200\,ns voltage pulse is applied to the device. The time-dependent current through the p-MTJ is tracked in real-time on a 50-Ohm terminated digital oscilloscope (Tektronix TDS 6604, with a pair of broad-band amplifies in front). A linear background is subtracted using the trace of a low bias non-switching event.   Figure\,\ref{fig3}(a) shows the p-MTJ conductivity $\sigma$ for  single shot P to AP transitions when a -0.5\,V pulse is applied at two different applied field values. From Fig.\,\ref{fig2}(c), both of these fields should exhibit  a low probability of errors. As expected, we observe  the junction conductivity decrease within roughly 15\,ns in both cases,  indicating  successful switching.

From Fig.\,\ref{fig2}(c), application of a -0.65\,V pulse will likely produce a successful switch in zero field, but with the application of 1.25\,kOe there is a high chance of a switching error. Fig.\,\ref{fig3}(b) shows that the p-MTJ switches within 6\,ns at zero field. However, in a 1.25\,kOe field the signal does not show a clean switching behavior. Instead, we observe an oscillatory signal with an amplitude near the full switching amplitude observed in zero field, indicating there is large angle dynamics beyond switching of the p-MTJ layers. The final state of the p-MTJ upon withdrawal of the voltage pulse will be determined by the probability the intermediate states are occupied at the moment the pulse is removed.

Auto-oscillations of a single layer's magnetization driven by spin currents have been observed in spin torque oscillator devices\cite{Kiselev2003,Demidov2012,Safranski2017}. The characteristic frequency of these devices is typically close to the ferromagnetic resonance frequency\cite{Slavin2009}. From previous measurements of similar p-MTJ made from the same material stack\citep{Safranski2016}, this frequency should be at a minimum 5\,GHz. However, the frequency of the large amplitude events observed here is less than  100\,MHz, suggesting that these observed dynamics are not the excitation of small angle auto-oscillations of the FL ferromagnetic resonance mode.

\begin{figure*}[tbp!]
\includegraphics[width=\textwidth]{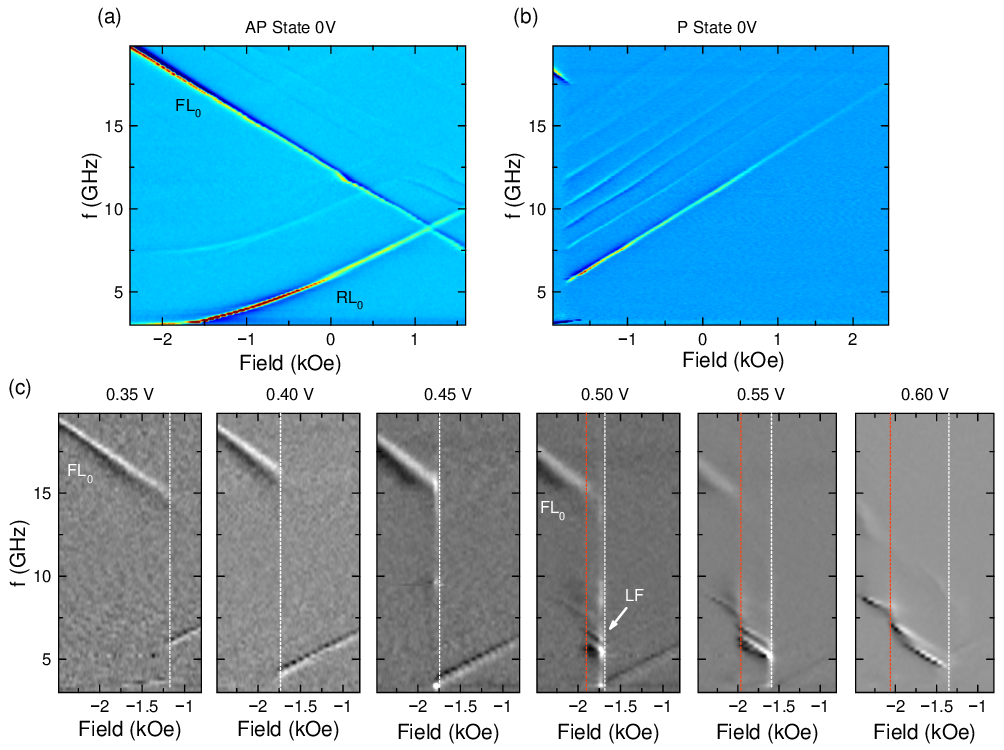}
\caption{
ST-FMR measurements showing the field dispersions in the (a) AP and (b) P states. (c) Field derivative of the  emitted microwave power as a function of applied field and detected frequency for multiple amplitude of voltage pulses. White line indicates switching boundary into the P state and red line shows the region where FL frequency drops. 
}
\label{fig_emission}
\end{figure*}

\subsection{Frequency Domain Characteristics}

Next we further investigate the frequency domain characteristics  when the p-MTJ is held in the oscillatory state. Fig.\,\ref{fig3}(c) shows the time domain conductivity when a 0.55\,V pulse is applied along   with a -1.75\,kOe  field, representing errors in an AP-to-P switching event. As seen before in Fig.\,\ref{fig3}(b), large oscillations in the junction conductivity are observed. We perform a discreet sliding  Fourier transformation on this trace with  a 7.5\,ns window over the duration of the pulse and assemble it into the color plot shown in Fig.\,\ref{fig3}(c), with the average of  these traces  shown on the right. The low frequency content in the spectra is consistent with the periodicity of the large amplitude oscillations. More interestingly, a peak near 5.5\,GHz is observed.  Given the large angle low frequency dynamics present, it is surprising to see a distinct  peak in this frequency range. The color plot also indicates that this frequency content is present over almost the entire length of the trace except for a brief period starting at around 120\,ns when the p-MTJ temporarily stabilizes in the AP state.  

Generally, microwave  frequency generation in p-MTJ can be interpreted as the excitation of the  standing spin wave modes in  ferromagnetic layers near the MgO interface\cite{Naletov2011,Goncalves2013,Safranski2016}. While the exact source of the signals produced is unclear in circular p-MTJ, the resonance field dispersion can be reliably obtained. To map out these modes, we use field modulated spin torque ferromagnetic resonance (ST-FMR) techniques\cite{Tulapurkar2005,Sankey2006,Goncalves2013}. We perform  measurements on the p-MTJ in the AP state and obtain the field derivative ST-FMR amplitude as a function of microwave drive frequency and applied magnetic field shown in Fig.\,\ref{fig_emission}(a).  Branches with two distinct slopes are seen and as determined in previous studies in similar devices\citep{Safranski2016}, resonance modes with positive slope are associated to the RL and those with negative slopes the FL. The lowest order modes $\mathrm{FL}_\mathrm{0}$ and $\mathrm{RL}_\mathrm{0}$ are the quasi-uniform modes, with higher frequency modes arising from the higher order standing spin wave modes. 

From Fig.\,\ref{fig_emission}(a) when the p-MTJ is in the initial AP state, there is no distinct source for the 5.5\,GHz peak seen in  the high bias time domain measurements. However, the conductivity shown in Fig.\,\ref{fig3}(c) suggests that the p-MTJ dwells in a state closer to the P state. Figure\,\ref{fig_emission}(b) shows the ST-FMR  field dispersion in the P state. Near -1.75\,kOe there is a resonance around  5.5\,GHz, but ascribing the peak to any of these resonances at low bias is difficult since joule heating and field like torques can drastically alter the resonance position\cite{Petit2007,Fu2016}. 

To investigate the high bias behavior of the p-MTJ spectra, we switch to microwave emission based measurements. The p-MTJ is supplied with 100\,ns voltage pulses with a repetition period of 1\,$\mu$s to reduce   Ohmic heating.  A spectrum analyzer is  used to measure microwave power produced by the sample.  To enhance the signal to noise in this measurement, we employ a field modulated spectrum analyzer method discussed elsewhere\cite{Safranski2017}. 

Figure\,\ref{fig_emission}(c) shows the field derivative of the microwave signal as a function of frequency and applied field for multiple amplitude voltage pulses. When a 0.35\,V pulse amplitude is used, we observe a negative sloping line left of the white dotted line where the p-MTJ is in the AP state.   From the ST-FMR measurements performed earlier, this is the lowest order spin wave mode $\mathrm{FL}_\mathrm{0}$. Right of the white dotted line, the p-MTJ has switched to the P state and we observe one of the P state resonance modes. Further increasing the pulse amplitude to 0.4\,V and 0.45\,V results in little change to the spectral content. The p-MTJ switches to the P state at larger fields (white line), consistent with Fig.\,\ref{fig2}(b). We also observe a slight decrease of the resonance frequency at a given field likely related to increased joule heating and subsequent decrease in  perpendicular anisotropy from temperature\cite{Fu2016}. 

When a 0.5\,V and larger pulse is applied, we observe a change in the spectral content. For larger negative fields we still observe the $\mathrm{FL}_\mathrm{0}$ resonance, but as the field is decreased, there is a sudden jump in frequency (red dotted line) and a new lower frequency mode (LF) is observed. The LF mode  has a negative slope like the FL modes in the AP state,  opposite of the P state resonance mode slopes. This indicates  the LF mode is not just the excitation of one of the P state modes even though Fig.\,\ref{fig3}(c) suggests the p-MTJ dwells closer to the P state. It  is also important to note that the normal $\mathrm{FL}_\mathrm{0}$ mode identified in ST-FMR measurements no longer exists when this new mode is observed. These observations suggest  that this mode is related to the FL dynamics.  The field and frequency of the peak observed in Fig.\,\ref{fig3}(c)  match up with the LF mode. While in the time domain based measurement there is only one main peak, there are multiple in the data obtained by the spectrum analyzer. Mode hopping events like those observed in spin torque oscillator devices\cite{Muduli2012}  could explain the multiple peaks in this data, as it integrates over a significantly longer time than the single time trace.

The field range the LF mode begins is similar to the range where high voltage switching errors are observed in Fig.\,\ref{fig2}(b) (white points), suggesting that the presence of the LF mode is related to such high-bias WER rise. An understanding of the onset of this low-frequency mode is therefore important to the reduction and removal of high-bias WER rise, which is important for STT-MRAM applications\cite{2019009}. 

The high-bias WER-rise has been reported before, and has been suggested to arise from a decrease of FL uniaxial anisotropy, presumably related to ohmic heating\cite{Sun2009,Strelkov2018}. The decrease in the FL resonance frequency with increasing voltage in Fig.\,\ref{fig_emission}(b) is consistent with this assertion. However, the jump in frequency at high voltages from the FMR-like mode to the LF mode is not, as a change in anisotropy with heating would be expected to be a smooth continuous function of bias. Effects such as voltage controlled magnetic anisotropy\cite{Ando2009} will alter  the resonance frequency and exchange stiffness\cite{Dohi2017}, but are known to do so continuously as well. Low frequency sub-harmonics have been observed in the non-linear regime\cite{Auerbach2018}, but do not correspond to the disappearance of the main resonance mode as seen in our measurements.

Next we explore the likely causes of this LF-mode, and identify possible mechanisms that give rise to a threshold-like onset for this mode {\it vs} bias voltage. We do so by numerically simulating the dynamics of the p-MTJ stack, and by going beyond the individual layer's macrospin limit, but only introduce the minimum amount of additional degrees-of-freedom as necessary. Turns out it is very difficult if not impossible for a simple three-moment stack to possess such characteristics of the LF-mode and simultaneously the type of WER-boundary as seen in Fig.\,\ref{fig2}(b) and (c). On the other hand,  we will next demonstrate that these WER-rise boundaries as shown in Fig.\,\ref{fig2}(b) and (c), together with the presence of a high-bias LF-mode strongly suggest the presence of interface moments with low magnetization, high local anisotropy, and limited exchange coupling to the rest of the layers, which is consistent with other behaviors we have seen from these p-MTJs.\cite{2015095,2019010,Sun2017}

\section{Numerical Simulations}

First in the following sections, we consider the simplest model of the p-MTJ, where each layer shown in Fig.\,\ref{fig1} is represented by a single macrospin. While the resulting three-moment model can be used to qualitatively explain the large angle dynamics seen in the time domain measurements, analysis of the frequency content produced by this model can not reproduce the observation of the low frequency microwave generation.  We then modify the macrospin model  to take into account interface moments represented by separate exchange coupled macrospins. This model shows  similar large angle dynamics to the three moment model, but more importantly is able to qualitatively capture the drop in resonance frequency observed in experiments.

\subsection{Three Macrospin-Moment Coupled Model}

\begin{figure}[pt]
\includegraphics[width=0.5\textwidth]{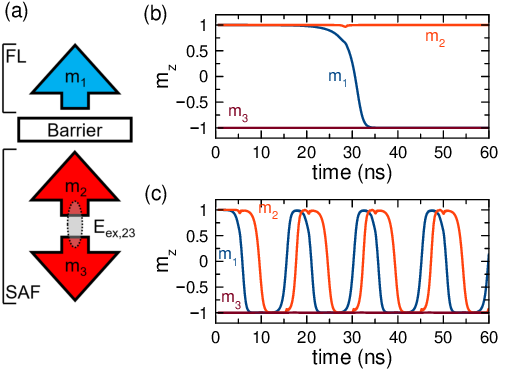}
\caption{(a) A macrospin construct  of the magnetic tunnel junction layers.  $z$ component of the magnetization of the layers near the interface for a (b) $5.3\times 10^4$\,$emu\,s^{-1}\,cm^{-2}$ and (c) $1.3\times10^5$\,$emu\,s^{-1}\,cm^{-2}$ pulses.  }
\label{fig_sim1}
\end{figure}

Here we build a numerical three moment coupled macrospin  model by writing their corresponding Landau-Lifshitz-Gilbert equations for a structure schematically shown in Fig.\,\ref{fig_sim1}(a). In this model we  consider the FL, RL and bottom SAF layers.  For simplicity, we assume the Slonczewski-type of spin-torque is only present at the tunnel interface between the FL moment $m_1$  and RL moment $m_2$. The RL $m_2$ is exchange coupled to  the bottom SAF layer $m_3$ with energy $E_\mathrm{ex,23} = -0.1\,emu\,cm^{-2}$.  The model then is then written
\begin{widetext}
\begin{equation}
\left\{
\begin{array}{l}
\dfrac{d\mathbf{n}_{m1}}{dt}=\gamma \left[  H_\mathrm{k1}\left( \mathbf{n}_\mathrm{k1}\cdot \mathbf{n}_\mathrm{m1}\right)\mathbf{n}_\mathrm{k1} +\mathbf{H}_\mathrm{a} \right]
   \times \mathbf{n}_{m1} +\alpha _1  \mathbf{n}_{m1} \times \dfrac{d\mathbf{n}_{m1}}{dt} +\left( \dfrac{J_\text{s}}{M_{s1}t_1}\right) \mathbf{n}_{m1} \times \left(\mathbf{n}_{m1}\times \mathbf{n}_{m2} \right) \\
\\
\dfrac{d\mathbf{n}_{m2}}{dt}=\gamma  \left[  H_\mathrm{k2}\left( \mathbf{n}_\mathrm{k2}\cdot \mathbf{n}_\mathrm{m2}\right)\mathbf{n}_\mathrm{k2} + \dfrac{E_\mathrm{ex,23}}{M_{s2}t_2} \mathbf{n}_\mathrm{m3}+\mathbf{H}_\mathrm{a} \right]
   \times \mathbf{n}_{m2} +\alpha _2  \mathbf{n}_{m2} \times \dfrac{d\mathbf{n}_{m2}}{dt} -\left( \dfrac{J_\text{s}}{M_{s2}t_2}\right) \mathbf{n}_{m2} \times \left(\mathbf{n}_{m2}\times \mathbf{n}_{m1} \right) \\
\\
\dfrac{d\mathbf{n}_{m3}}{dt}=\gamma \left[  H_\mathrm{k3}\left( \mathbf{n}_\mathrm{k3}\cdot \mathbf{n}_\mathrm{m3}\right)\mathbf{n}_\mathrm{k3} + \dfrac{E_\mathrm{ex,23}}{M_{s3}t_3}\mathbf{n}_\mathrm{m2}+\mathbf{H}_\mathrm{a} \right]
   \times \mathbf{n}_{m3} +\alpha _3 \mathbf{n}_{m3} \times \dfrac{d\mathbf{n}_{m3}}{dt}\\
\end{array}
\right.\label{E1}
\end{equation}
\end{widetext}
where $\mathbf{n}_\mathrm{m1,2,3}$ are the moment direction unit vectors, $H_\mathrm{k1,2,3}$ are the uniaxial anisotropy fields of the individual moments, $\mathbf{n}_\mathrm{k1,2,3}=\mathbf{e}_\mathrm{z}$ are their collinear anisotropy axes (along the z-direction or film-normal for PMA); $\alpha_\mathrm{1,2,3}$ are the LLG-damping factor for the moments; $J_\mathrm{s}$ is the spin current flowing through the layer in units of magnetic moment, generating spin-torque across the interface between $m_1$ and $m_2$. For simplicity, we assume equal charge to spin conversion efficiency for both layers.    $\gamma \approx 2 \mu_B/\hbar$ is the gyro-magnetic ratio, and $\mathbf{H}_a$ is the applied field, along $\mathbf{e}_z$ also. The values used for these parameters can be seen in Table\,\ref{sim1_table}.


\begin{table}[]
\begin{tabular}{ccccc}
   \hline
   \hline
   Moment & $M_{\mathrm{s}}$ ($emu\,cm^{-3}$)  & $H_{k}$ (kOe)    & $\alpha$ & t (nm)   \\
   \hline
$m_1$ & 700 & 4   & 0.005 & 1.5 \\ 
$m_2$ & 700 & 2 & 0.01 & 1.5 \\
$m_3$ & 700 & 6 & 0.02  & 2 \\
\hline
\hline
\end{tabular}
\caption{Material parameters used in the three macrospin-moment model .}
\label{sim1_table}
\end{table}

Figure\,\ref{fig_sim1}(b) shows the z component of  $m_1$, $m_2$, and $m_3$  as a function of time when a $J_\mathrm{s}=5.3\times 10^4$\,$emu\,s^{-1}\,cm^{-2}$ amplitude pulse is applied. The system is initially prepared in the P state and as time passes,  $m_1$ flips with no significant change in $m_2$ representing a successful switching event.  Figure\,\ref{fig_sim1}(c) shows the behavior when the pulse amplitude is increased to $1.3\times10^5$\,$emu\,s^{-1}\,cm^{-2}$. Shortly after  $m_1$ flips, $m_2$ flips its direction  and a ``pinwheel'' process\cite{Kim2008} begins.   Qualitatively, this is  similar to the large amplitude oscillations  observed in experiment. 


These numerical simulations can relate the large oscillatory signal seen in experiment at high bias to the pinwheeling between the FL and RL layers. However, it can not recreate the switching characteristics seen with varying the applied magnetic field in experiment. Mainly, for positive applied magnetic fields along the RL direction, the field stabilizes the RL and prevents the pinwheel state from forming. This is contrary to the observation in Fig.\,\ref{fig2}(c) where the WER-rise  threshold  decreases as magnetic field is swept in the positive direction. Further, we perform FFT analysis on the  time domain dynamics and observe no resonance peak in this state, indicating the model fails to explain the LF mode. This mismatch with the experimental observations suggests that the model needs to be expanded to more accurately represent the layer material properties.

\subsection{Five Macrospin-Moment Coupled  Model}

The presence of the LF mode points to magnetization dynamics outside of the simplest mathematical pictures of p-MTJ, where each representative layer is taken to be uniform.  In actual materials,  perpendicular  anisotropy  is expected to be  largest at atomic layers near  interfaces\cite{Carcia1985,Hallal2013,Cuadrado2018}. Experimental studies have shown that the MgO/FM interface can  form magnetic oxides\cite{Barsukov2015,Fu2016}, which will result in a large variation in the interface moment properties across the width of the layer. Taking these aspects into consideration, models treating the interface and bulk separately have been able to describe the size and RA dependence of the switching threshold in p-MTJ\cite{Sun2017}. 

\begin{table}[]
\begin{tabular}{ccccc}
   \hline
   \hline
   Moment & $M_{\mathrm{s}}$ ($emu\,cm^{-3}$)  & $H_{k}$ (kOe)    & $\alpha$ & t (nm)   \\
   \hline
$m_1$ & 600 & 1   & 0.007 & 1.5 \\ 
$m_2$ & 160 & 21 & 0.005 & 0.4 \\
$m_3$ & 175 & 22 & 0.012  & 0.4 \\
$m_4$ & 1200 & 2  & 0.012  & 3   \\
$m_5$ & 1200 & 2  & 0.02  & 5  \\
\hline
\hline
\end{tabular}
\caption{Material parameters used in the five macrospin-moment model where interfaces are taken into account. }
\label{sim2_table}
\end{table}

\begin{figure*}[tbp!]
\includegraphics[width=\textwidth]{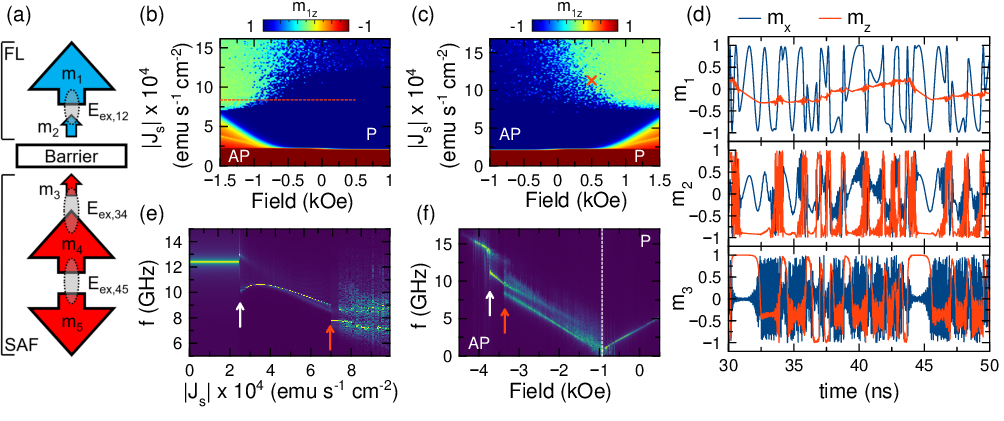}
\caption{
(a) A macrospin-construct of the magnetic tunnel junction taking into account the interface moments  Spin-torque exists only across the interface between $m_2$ and $m_3$. Field-bias switching/error contour, where the amplitude average of $m_{1z}$ is used as a switching error proxy   for crude qualitative comparison with experiment for the (b) AP-to-P and (c) P-to-AP transitions. Red line in (b) corresponds to bias used in (f). Red x in (c) indicates field (500\,Oe) and bias ($1.1\times10^5$\,$emu\,s^{-1}\,cm^{-2}$) used in (d) the simulated time-trace for z-components (red) and x-components (blue) of the top three layers. (e) Resonance frequency of $m_1$ as a function of bias at -3\,kOe of field -- corresponding to a condition that is outside to the left of panel-(b). White arrow indicates onset of $m_2$ instability and red arrow the onset of pinwheeling. (f) $m_1$ frequency vs field with $-8.5\times10^4$\,$emu\,s^{-1}\,cm^{-2}$ of bias. White arrow represents saturation of $m_3$ moment  and red arrow the onset of pinwheeling. 
}
\label{fig_sim2}
\end{figure*}

We therefore modify the numerical model shown above to treat the dynamics of tunnel-barrier interface-bound magnetic moments separately, coupled to the rest of the film stack. We further assume that spin-current transmission and absorption across the MgO tunnel barrier is fully accounted for at these two interface layers. Such a five-coupled-moment model is schematically represented  in Fig.\,\ref{fig_sim2}(a). The FL bulk is represented by $m_1$, the FL interface by $m_2$. Similarly, the top RL interface is represented by $m_3$, and the rest of top RL by $m_4$. $m_5$ represents RL2. With a negative exchange coupling $E_\mathrm{ex,45}$, RL1 and RL2 together describes a SAF reference layer construct. 

We then write out the corresponding LLG equations,  similar to those in eq.\,\ref{E1} using the parameters shown in Table\,\ref{sim2_table}, and with $E_\mathrm{ex,12} = 0.06\,emu\,cm^{-2}$, $E_\mathrm{ex,34} = 0.1\,emu\,cm^{-2}$ and $E_\mathrm{ex,45} = -0.4\,emu\,cm^{-2}$.  These materials-related parameters are chosen to be consistent with literature values associated with interface-related magnetic texture discussion\cite{Hallal2013}, and with our observable device properties. They are however only for qualitative numerical illustration of the mechanisms involved, as our device structures details may be significantly more complex. Numerically evaluated model results are compared with experimentally obtained time-dependent junction conductance trances and corresponding frequency-domain spectra, as well as with WER-rise boundary in $V_\mathrm{pulse}$ and $\mathbf{H}_a$, such as those shown in Fig.\,\ref{fig2}.

To keep numerical procedures simple and avoid computation time and resources for a probabilistic calculation, we use the relevant moment's z-component value obtained from time domain dynamics to approximate the switching and reversion-error regions in the $\left(J_s, \mathbf{H}_a \right)$ parameter space. Specifically, the average of the z component of the $m_1$ magnetization  from 33\,ns to 50\,ns is used as a proxy for estimating the boundaries of switching and/or error-generation. 

Figure\,\ref{fig_sim2}(b) shows the resulting $m_{1z}$ for the AP-to-P transition where $m_{1z}$ is initially set to -1. As seen in experiments, the main switching threshold, as represented by the lower $\left| J_\mathrm{s}\right|$ boundary in Figure\,\ref{fig_sim2}(b) depends on $H_\mathrm{a}$.  For larger values of $\left|J_\mathrm{s}\right|$, a second region emerges where there is no longer a full reversal of $m_{1z}$, representing the WER-rise region. The $V_\mathrm{pulse} - \mathbf{H}_a$ dependent  WER-rise boundary shows stronger field dependence than the main switching threshold's. The same is seen in both AP-P and P-AP STT-switching directions, as shown in Fig.\,\ref{fig_sim2}(b) and (c), with the FL's orientation with respect to applied field $\mathbf{H}_a$ dictating the sign of the WER-rise boundary slope.

Figure\,\ref{fig_sim2}(d) shows the time domain dynamics for the top three layers in the model when a $J_\mathrm{s} = 1.1\times 10^5$\,$emu\,s^{-1}\,cm^{-2}$ is applied with a 500\,Oe field, corresponding to a region where high bias switching errors appear as shown by the red x in Fig.\,\ref{fig_sim2}(c). We observe that a dynamic state is formed between the interface moments $m_2$ and $m_3$, similar to the ``pinwheel'' motion in the three moment model, albeit now at frequencies much higher than that of the FL or RL bulk moment's.  

Once the two interface moments enter their large amplitude precessions, their exchange-coupled contribution to the perpendicular anisotropy of the bulk of FL $m_1$ (and RL $m_4$) is diminished. This is revealed by the emergence of a new, low-frequency (LF) spectral signature of the FL-bulk $m_1$ at large values $\left|J_\mathrm{s}\right|$ in the WER-rise region in Fig.\,\ref{fig_sim2}(b-c), as illustrated in Figure\,\ref{fig_sim2}(e-f), which we will discuss below.

If the threshold for the WER-rise onset in Figure\,\ref{fig_sim2}(b-c) is due to the loss of PMA from interface moment's developing their large amplitude dynamics, then we expect to see a drop in the $m_1$ resonance frequency when a bias and field value inside the WER-rise region is used.  Figure\,\ref{fig_sim2}(e) shows the precession frequency of the $m_{1x}$ component as a function of applied bias with -3\,kOe of magnetic field, obtained from FFT of the time domain dynamics.  At low bias, $m_1$ and $m_2$ precess in sync at a frequency corresponding to the effective anisotropy of the  pair and applied field. When a sufficient drive is applied, the $m_2$ interface moment enters a large angle state and a sudden drop in the $m_1$ resonance frequency is observed  (white arrow). Further increasing the bias leads to a pinwheeling state (red arrow) between the $m_2$ and $m_3$ moments. 

To further illustrate the process associated with Fig.\,\ref{fig_emission}(c), we simulate the  field dispersion when $-8.5\times10^4$\,$emu\,s^{-1}\,cm^{-2}$ of bias is applied, shown in Fig.\,\ref{fig_sim2}(f), corresponding to the red dashed line drawn in Fig.\,\ref{fig_emission}(b).  At large negative fields, the RL interface moment is saturated in the negative direction resulting in no excitation of the FL. The $m_1$ and $m_2$ pair then relax to equilibrium at their combined resonance frequency. Once the field is lowered, the RL interface moments remain in their  initially defined positive direction, allowing an antidamping torque to be exerted on the FL. The $m_2$ moments then enter a large angle state and the resonance frequency of $m_1$ drops(white arrow). Further decreasing the magnetic field, there is a second jump (red arrow) at the  point the RL interface $m_3$ is no longer stable and enters a pinwheel state with $m_2$,  similar to that in Fig.\,\ref{fig_sim2}(d). Around -1\,kOe (white dotted line) the device successfully switches into the P state as shown in    Fig.\,\ref{fig_emission}(b). At this point the large angle dynamics are no longer excited, and the $m_1$ and $m_2$ pair precess at their combined resonance frequency once more. The numerical simulation behavior is qualitatively similar to the measurements in Fig\,\ref{fig_emission}(c), where at high biases multiple jumps in frequency are observed.

These numerical studies above demonstrate and confirm the understanding that high-frequency interface-moment dynamics play a important role in switching and error-generation behavior in our p-MTJs. The large-amplitude excitation of interface moments with dynamic frequencies well-above the bulk of FL or RL causes the effective perpendicular anisotropy of the bulk FL and RL to decrease. This reduction to the energy barriers causes an increase of the error-probability due to thermal agitation and disturbances from coupled interface moment dynamics. This type of dynamic reduction of effective energy barrier would cause the observed WER-rise at higher voltages\cite{2019009}. 

\section{Conclusion}

We demonstrate the important role interface moments play in STT-driven switching dynamics for our p-MTJs. Experiments  at high bias voltages reveal a WER-rise (i.e. switch-reversion) behavior. Correspondingly we observe the onset of a new LF mode in the spectra-content of the real-time p-MTJ conductance at high-bias. A five macrospin coupled numerical model is successful in illustrating the important role interface moments play, when such interface moments are situated in large perpendicular interface anisotropy and with limited exchange-coupling to the bulk of corresponding FL and RL. Their high-frequency dynamic excitation under STT is shown to be responsible for the resulting LF mode, as well as the observed WER-rise state-boundary in $\left( J_\mathrm{s}, \mathbf{H}_a \right)$ space.

\section*{Acknowledgment}
We acknowledge fruitful discussions with our colleagues; in particular, with Janusz Nowak for WER properties and mechanisms related to our p-MTJs, with Guohan Hu and Matthias Gottwald for materials properties concerning our understanding of an tunnel barrier interface concentrated perpendicular anisotropy. Work done with the MRAM group at IBM T. J. Watson Research Center in Yorktown Heights, New York, and is supported in part by partnership with Samsung Electronics. 


\begin{thebibliography}{43}
\expandafter\ifx\csname natexlab\endcsname\relax\def\natexlab#1{#1}\fi
\expandafter\ifx\csname bibnamefont\endcsname\relax
  \def\bibnamefont#1{#1}\fi
\expandafter\ifx\csname bibfnamefont\endcsname\relax
  \def\bibfnamefont#1{#1}\fi
\expandafter\ifx\csname citenamefont\endcsname\relax
  \def\citenamefont#1{#1}\fi
\expandafter\ifx\csname url\endcsname\relax
  \def\url#1{\texttt{#1}}\fi
\expandafter\ifx\csname urlprefix\endcsname\relax\def\urlprefix{URL }\fi
\providecommand{\bibinfo}[2]{#2}
\providecommand{\eprint}[2][]{\url{#2}}

\bibitem[{\citenamefont{Slonczewski}(1996)}]{Slonczewski1996}
\bibinfo{author}{\bibfnamefont{J.}~\bibnamefont{Slonczewski}},
  \bibinfo{journal}{J. Magn. Magn. Mater.} \textbf{\bibinfo{volume}{159}},
  \bibinfo{pages}{L1} (\bibinfo{year}{1996}).

\bibitem[{\citenamefont{Berger}(1996)}]{Berger1996}
\bibinfo{author}{\bibfnamefont{L.}~\bibnamefont{Berger}},
  \bibinfo{journal}{Phys. Rev. B} \textbf{\bibinfo{volume}{54}},
  \bibinfo{pages}{9353} (\bibinfo{year}{1996}).

\bibitem[{\citenamefont{Sun}(1999)}]{Sun1999}
\bibinfo{author}{\bibfnamefont{J.~Z.} \bibnamefont{Sun}}, \bibinfo{journal}{J.
  Magn. Magn. Mater.} \textbf{\bibinfo{volume}{202}}, \bibinfo{pages}{157}
  (\bibinfo{year}{1999}).

\bibitem[{\citenamefont{Worledge et~al.}(2011)\citenamefont{Worledge, Hu,
  Abraham, Sun, Trouilloud, Nowak, Brown, Gaidis, O'Sullivan, and
  Robertazzi}}]{Worledge2011}
\bibinfo{author}{\bibfnamefont{D.~C.} \bibnamefont{Worledge}},
  \bibinfo{author}{\bibfnamefont{G.}~\bibnamefont{Hu}},
  \bibinfo{author}{\bibfnamefont{D.~W.} \bibnamefont{Abraham}},
  \bibinfo{author}{\bibfnamefont{J.~Z.} \bibnamefont{Sun}},
  \bibinfo{author}{\bibfnamefont{P.~L.} \bibnamefont{Trouilloud}},
  \bibinfo{author}{\bibfnamefont{J.}~\bibnamefont{Nowak}},
  \bibinfo{author}{\bibfnamefont{S.}~\bibnamefont{Brown}},
  \bibinfo{author}{\bibfnamefont{M.~C.} \bibnamefont{Gaidis}},
  \bibinfo{author}{\bibfnamefont{E.~J.} \bibnamefont{O'Sullivan}},
  \bibnamefont{and} \bibinfo{author}{\bibfnamefont{R.~P.}
  \bibnamefont{Robertazzi}}, \bibinfo{journal}{Appl. Phys. Lett.}
  \textbf{\bibinfo{volume}{98}}, \bibinfo{pages}{98} (\bibinfo{year}{2011}).

\bibitem[{\citenamefont{Ikeda et~al.}(2010)\citenamefont{Ikeda, Miura,
  Yamamoto, Mizunuma, Gan, Endo, Kanai, Hayakawa, Matsukura, and
  Ohno}}]{Ikeda2010}
\bibinfo{author}{\bibfnamefont{S.}~\bibnamefont{Ikeda}},
  \bibinfo{author}{\bibfnamefont{K.}~\bibnamefont{Miura}},
  \bibinfo{author}{\bibfnamefont{H.}~\bibnamefont{Yamamoto}},
  \bibinfo{author}{\bibfnamefont{K.}~\bibnamefont{Mizunuma}},
  \bibinfo{author}{\bibfnamefont{H.~D.} \bibnamefont{Gan}},
  \bibinfo{author}{\bibfnamefont{M.}~\bibnamefont{Endo}},
  \bibinfo{author}{\bibfnamefont{S.}~\bibnamefont{Kanai}},
  \bibinfo{author}{\bibfnamefont{J.}~\bibnamefont{Hayakawa}},
  \bibinfo{author}{\bibfnamefont{F.}~\bibnamefont{Matsukura}},
  \bibnamefont{and} \bibinfo{author}{\bibfnamefont{H.}~\bibnamefont{Ohno}},
  \bibinfo{journal}{Nat. Mater.} \textbf{\bibinfo{volume}{9}},
  \bibinfo{pages}{721} (\bibinfo{year}{2010}).

\bibitem[{\citenamefont{Apalkov et~al.}(2016)\citenamefont{Apalkov, Dieny, and
  Slaughter}}]{Apalkov2016}
\bibinfo{author}{\bibfnamefont{D.}~\bibnamefont{Apalkov}},
  \bibinfo{author}{\bibfnamefont{B.}~\bibnamefont{Dieny}}, \bibnamefont{and}
  \bibinfo{author}{\bibfnamefont{J.~M.} \bibnamefont{Slaughter}},
  \bibinfo{journal}{Proc. IEEE} \textbf{\bibinfo{volume}{104}},
  \bibinfo{pages}{1796} (\bibinfo{year}{2016}).

\bibitem[{\citenamefont{Kent and Worledge}(2015)}]{2017019}
\bibinfo{author}{\bibfnamefont{A.~D.} \bibnamefont{Kent}} \bibnamefont{and}
  \bibinfo{author}{\bibfnamefont{D.~C.} \bibnamefont{Worledge}},
  \bibinfo{journal}{Nat. Nanotechnol.} \textbf{\bibinfo{volume}{10}},
  \bibinfo{pages}{187} (\bibinfo{year}{2015}).

\bibitem[{\citenamefont{Torrejon et~al.}(2017)\citenamefont{Torrejon, Riou,
  Araujo, Tsunegi, Khalsa, Querlioz, Bortolotti, Cros, Yakushiji, Fukushima
  et~al.}}]{Torrejon2017}
\bibinfo{author}{\bibfnamefont{J.}~\bibnamefont{Torrejon}},
  \bibinfo{author}{\bibfnamefont{M.}~\bibnamefont{Riou}},
  \bibinfo{author}{\bibfnamefont{F.~A.} \bibnamefont{Araujo}},
  \bibinfo{author}{\bibfnamefont{S.}~\bibnamefont{Tsunegi}},
  \bibinfo{author}{\bibfnamefont{G.}~\bibnamefont{Khalsa}},
  \bibinfo{author}{\bibfnamefont{D.}~\bibnamefont{Querlioz}},
  \bibinfo{author}{\bibfnamefont{P.}~\bibnamefont{Bortolotti}},
  \bibinfo{author}{\bibfnamefont{V.}~\bibnamefont{Cros}},
  \bibinfo{author}{\bibfnamefont{K.}~\bibnamefont{Yakushiji}},
  \bibinfo{author}{\bibfnamefont{A.}~\bibnamefont{Fukushima}},
  \bibnamefont{et~al.}, \bibinfo{journal}{Nature}
  \textbf{\bibinfo{volume}{547}}, \bibinfo{pages}{428} (\bibinfo{year}{2017}).

\bibitem[{\citenamefont{Camsari et~al.}(2019)\citenamefont{Camsari, Sutton, and
  Datta}}]{Camsari2018}
\bibinfo{author}{\bibfnamefont{K.~Y.} \bibnamefont{Camsari}},
  \bibinfo{author}{\bibfnamefont{B.~M.} \bibnamefont{Sutton}},
  \bibnamefont{and} \bibinfo{author}{\bibfnamefont{S.}~\bibnamefont{Datta}},
  \bibinfo{journal}{Appl. Phys. Rev.} \textbf{\bibinfo{volume}{6}},
  \bibinfo{pages}{011305} (\bibinfo{year}{2019}).

\bibitem[{\citenamefont{Nowak et~al.}(2011)\citenamefont{Nowak, Robertazzi,
  Sun, Hu, Abraham, Trouilloud, Brown, Gaidis, O'Sullivan, Gallagher
  et~al.}}]{Nowak2011}
\bibinfo{author}{\bibfnamefont{J.~J.} \bibnamefont{Nowak}},
  \bibinfo{author}{\bibfnamefont{R.~P.} \bibnamefont{Robertazzi}},
  \bibinfo{author}{\bibfnamefont{J.~Z.} \bibnamefont{Sun}},
  \bibinfo{author}{\bibfnamefont{G.}~\bibnamefont{Hu}},
  \bibinfo{author}{\bibfnamefont{D.~W.} \bibnamefont{Abraham}},
  \bibinfo{author}{\bibfnamefont{P.~L.} \bibnamefont{Trouilloud}},
  \bibinfo{author}{\bibfnamefont{S.}~\bibnamefont{Brown}},
  \bibinfo{author}{\bibfnamefont{M.~C.} \bibnamefont{Gaidis}},
  \bibinfo{author}{\bibfnamefont{E.~J.} \bibnamefont{O'Sullivan}},
  \bibinfo{author}{\bibfnamefont{W.~J.} \bibnamefont{Gallagher}},
  \bibnamefont{et~al.}, \bibinfo{journal}{IEEE Magn. Lett.}
  \textbf{\bibinfo{volume}{2}} (\bibinfo{year}{2011}).

\bibitem[{\citenamefont{Nowak et~al.}(2016)\citenamefont{Nowak, Robertazzi,
  Sun, Hu, Park, Lee, Annunziata, Lauer, Kothandaraman, O'Sullivan
  et~al.}}]{2016085}
\bibinfo{author}{\bibfnamefont{J.~J.} \bibnamefont{Nowak}},
  \bibinfo{author}{\bibfnamefont{R.~P.} \bibnamefont{Robertazzi}},
  \bibinfo{author}{\bibfnamefont{J.~Z.} \bibnamefont{Sun}},
  \bibinfo{author}{\bibfnamefont{G.}~\bibnamefont{Hu}},
  \bibinfo{author}{\bibfnamefont{J.~H.} \bibnamefont{Park}},
  \bibinfo{author}{\bibfnamefont{J.~H.} \bibnamefont{Lee}},
  \bibinfo{author}{\bibfnamefont{A.~J.} \bibnamefont{Annunziata}},
  \bibinfo{author}{\bibfnamefont{G.~P.} \bibnamefont{Lauer}},
  \bibinfo{author}{\bibfnamefont{C.}~\bibnamefont{Kothandaraman}},
  \bibinfo{author}{\bibfnamefont{E.~J.} \bibnamefont{O'Sullivan}},
  \bibnamefont{et~al.}, \bibinfo{journal}{IEEE Magn. Lett.}
  \textbf{\bibinfo{volume}{7}}, \bibinfo{pages}{3102604}
  (\bibinfo{year}{2016}).

\bibitem[{\citenamefont{Slonczewski}(2005)}]{2005129}
\bibinfo{author}{\bibfnamefont{J.~C.} \bibnamefont{Slonczewski}},
  \bibinfo{journal}{Phys. Rev. B} \textbf{\bibinfo{volume}{71}},
  \bibinfo{pages}{024411} (\bibinfo{year}{2005}).

\bibitem[{\citenamefont{Slonczewski}(1989)}]{2005073}
\bibinfo{author}{\bibfnamefont{J.~C.} \bibnamefont{Slonczewski}},
  \bibinfo{journal}{Phys. Rev. B} \textbf{\bibinfo{volume}{39}},
  \bibinfo{pages}{6995} (\bibinfo{year}{1989}).

\bibitem[{\citenamefont{Kim et~al.}(2008)\citenamefont{Kim, Lee, and
  Lee}}]{Kim2008}
\bibinfo{author}{\bibfnamefont{W.}~\bibnamefont{Kim}},
  \bibinfo{author}{\bibfnamefont{T.~D.} \bibnamefont{Lee}}, \bibnamefont{and}
  \bibinfo{author}{\bibfnamefont{K.~J.} \bibnamefont{Lee}},
  \bibinfo{journal}{Appl. Phys. Lett.} \textbf{\bibinfo{volume}{93}},
  \bibinfo{pages}{2008} (\bibinfo{year}{2008}).

\bibitem[{\citenamefont{Sun et~al.}(2009)\citenamefont{Sun, Gaidis, Hu,
  O'Sullivan, Brown, Nowak, Trouilloud, and Worledge}}]{Sun2009}
\bibinfo{author}{\bibfnamefont{J.~Z.} \bibnamefont{Sun}},
  \bibinfo{author}{\bibfnamefont{M.~C.} \bibnamefont{Gaidis}},
  \bibinfo{author}{\bibfnamefont{G.}~\bibnamefont{Hu}},
  \bibinfo{author}{\bibfnamefont{E.~J.} \bibnamefont{O'Sullivan}},
  \bibinfo{author}{\bibfnamefont{S.~L.} \bibnamefont{Brown}},
  \bibinfo{author}{\bibfnamefont{J.~J.} \bibnamefont{Nowak}},
  \bibinfo{author}{\bibfnamefont{P.~L.} \bibnamefont{Trouilloud}},
  \bibnamefont{and} \bibinfo{author}{\bibfnamefont{D.~C.}
  \bibnamefont{Worledge}}, \bibinfo{journal}{J. Appl. Phys.}
  \textbf{\bibinfo{volume}{105}} (\bibinfo{year}{2009}).

\bibitem[{\citenamefont{Choi et~al.}(2016)\citenamefont{Choi, Katine, Mangin,
  and Fullerton}}]{Choi2016}
\bibinfo{author}{\bibfnamefont{R.}~\bibnamefont{Choi}},
  \bibinfo{author}{\bibfnamefont{J.~A.} \bibnamefont{Katine}},
  \bibinfo{author}{\bibfnamefont{S.}~\bibnamefont{Mangin}}, \bibnamefont{and}
  \bibinfo{author}{\bibfnamefont{E.~E.} \bibnamefont{Fullerton}},
  \bibinfo{journal}{IEEE Trans. Magn.} \textbf{\bibinfo{volume}{52}}
  (\bibinfo{year}{2016}).

\bibitem[{\citenamefont{Yoshida et~al.}(2018)\citenamefont{Yoshida, Tanaka,
  Ataka, and Furuya}}]{Yoshida2018}
\bibinfo{author}{\bibfnamefont{C.}~\bibnamefont{Yoshida}},
  \bibinfo{author}{\bibfnamefont{T.}~\bibnamefont{Tanaka}},
  \bibinfo{author}{\bibfnamefont{T.}~\bibnamefont{Ataka}}, \bibnamefont{and}
  \bibinfo{author}{\bibfnamefont{A.}~\bibnamefont{Furuya}},
  \bibinfo{journal}{2018 Non-Volatile Memory Technology Symposium (NVMTS)} pp.
  \bibinfo{pages}{1--5} (\bibinfo{year}{2018}).

\bibitem[{\citenamefont{Jan et~al.}(2018)\citenamefont{Jan, Thomas, Le, Lee,
  Liu, Zhu, Iwata-Harms, Patel, Tong, Sundar et~al.}}]{2019009}
\bibinfo{author}{\bibfnamefont{G.}~\bibnamefont{Jan}},
  \bibinfo{author}{\bibfnamefont{L.}~\bibnamefont{Thomas}},
  \bibinfo{author}{\bibfnamefont{S.}~\bibnamefont{Le}},
  \bibinfo{author}{\bibfnamefont{Y.-J.} \bibnamefont{Lee}},
  \bibinfo{author}{\bibfnamefont{H.}~\bibnamefont{Liu}},
  \bibinfo{author}{\bibfnamefont{J.}~\bibnamefont{Zhu}},
  \bibinfo{author}{\bibfnamefont{J.}~\bibnamefont{Iwata-Harms}},
  \bibinfo{author}{\bibfnamefont{S.}~\bibnamefont{Patel}},
  \bibinfo{author}{\bibfnamefont{R.-Y.} \bibnamefont{Tong}},
  \bibinfo{author}{\bibfnamefont{V.}~\bibnamefont{Sundar}},
  \bibnamefont{et~al.}, \bibinfo{journal}{VLSI Tech. Symp. IEEE}
  p.~\bibinfo{pages}{65} (\bibinfo{year}{2018}).

\bibitem[{\citenamefont{Cuadrado et~al.}(2018)\citenamefont{Cuadrado,
  Oroszl{\'{a}}ny, De{\'{a}}k, Ostler, Meo, Chepulskii, Apalkov, Evans,
  Szunyogh, and Chantrell}}]{Cuadrado2018}
\bibinfo{author}{\bibfnamefont{R.}~\bibnamefont{Cuadrado}},
  \bibinfo{author}{\bibfnamefont{L.}~\bibnamefont{Oroszl{\'{a}}ny}},
  \bibinfo{author}{\bibfnamefont{A.}~\bibnamefont{De{\'{a}}k}},
  \bibinfo{author}{\bibfnamefont{T.~A.} \bibnamefont{Ostler}},
  \bibinfo{author}{\bibfnamefont{A.}~\bibnamefont{Meo}},
  \bibinfo{author}{\bibfnamefont{R.~V.} \bibnamefont{Chepulskii}},
  \bibinfo{author}{\bibfnamefont{D.}~\bibnamefont{Apalkov}},
  \bibinfo{author}{\bibfnamefont{R.~F.} \bibnamefont{Evans}},
  \bibinfo{author}{\bibfnamefont{L.}~\bibnamefont{Szunyogh}}, \bibnamefont{and}
  \bibinfo{author}{\bibfnamefont{R.~W.} \bibnamefont{Chantrell}},
  \bibinfo{journal}{Phys. Rev. Appl.} \textbf{\bibinfo{volume}{9}},
  \bibinfo{pages}{54048} (\bibinfo{year}{2018}).

\bibitem[{\citenamefont{Hallal et~al.}(2013)\citenamefont{Hallal, Yang, Dieny,
  and Chshiev}}]{Hallal2013}
\bibinfo{author}{\bibfnamefont{A.}~\bibnamefont{Hallal}},
  \bibinfo{author}{\bibfnamefont{H.~X.} \bibnamefont{Yang}},
  \bibinfo{author}{\bibfnamefont{B.}~\bibnamefont{Dieny}}, \bibnamefont{and}
  \bibinfo{author}{\bibfnamefont{M.}~\bibnamefont{Chshiev}},
  \bibinfo{journal}{Phys. Rev. B} \textbf{\bibinfo{volume}{88}},
  \bibinfo{pages}{1} (\bibinfo{year}{2013}).

\bibitem[{\citenamefont{Barsukov et~al.}(2015)\citenamefont{Barsukov, Fu,
  Safranski, Chen, Youngblood, Gon{\c{c}}alves, Spasova, Farle, Katine, Kuo
  et~al.}}]{Barsukov2015}
\bibinfo{author}{\bibfnamefont{I.}~\bibnamefont{Barsukov}},
  \bibinfo{author}{\bibfnamefont{Y.}~\bibnamefont{Fu}},
  \bibinfo{author}{\bibfnamefont{C.}~\bibnamefont{Safranski}},
  \bibinfo{author}{\bibfnamefont{Y.~J.} \bibnamefont{Chen}},
  \bibinfo{author}{\bibfnamefont{B.}~\bibnamefont{Youngblood}},
  \bibinfo{author}{\bibfnamefont{A.~M.} \bibnamefont{Gon{\c{c}}alves}},
  \bibinfo{author}{\bibfnamefont{M.}~\bibnamefont{Spasova}},
  \bibinfo{author}{\bibfnamefont{M.}~\bibnamefont{Farle}},
  \bibinfo{author}{\bibfnamefont{J.~A.} \bibnamefont{Katine}},
  \bibinfo{author}{\bibfnamefont{C.~C.} \bibnamefont{Kuo}},
  \bibnamefont{et~al.}, \bibinfo{journal}{Appl. Phys. Lett.}
  \textbf{\bibinfo{volume}{106}} (\bibinfo{year}{2015}).

\bibitem[{\citenamefont{Fu et~al.}(2016)\citenamefont{Fu, Barsukov, Li,
  Gon{\c{c}}alves, Kuo, Farle, and Krivorotov}}]{Fu2016}
\bibinfo{author}{\bibfnamefont{Y.}~\bibnamefont{Fu}},
  \bibinfo{author}{\bibfnamefont{I.}~\bibnamefont{Barsukov}},
  \bibinfo{author}{\bibfnamefont{J.}~\bibnamefont{Li}},
  \bibinfo{author}{\bibfnamefont{A.~M.} \bibnamefont{Gon{\c{c}}alves}},
  \bibinfo{author}{\bibfnamefont{C.~C.} \bibnamefont{Kuo}},
  \bibinfo{author}{\bibfnamefont{M.}~\bibnamefont{Farle}}, \bibnamefont{and}
  \bibinfo{author}{\bibfnamefont{I.~N.} \bibnamefont{Krivorotov}},
  \bibinfo{journal}{Appl. Phys. Lett.} \textbf{\bibinfo{volume}{108}}
  (\bibinfo{year}{2016}).

\bibitem[{\citenamefont{Choi et~al.}(2007)\citenamefont{Choi, Tsunekawa,
  Nagamine, and Djayaprawira}}]{2007079}
\bibinfo{author}{\bibfnamefont{Y.~S.} \bibnamefont{Choi}},
  \bibinfo{author}{\bibfnamefont{K.}~\bibnamefont{Tsunekawa}},
  \bibinfo{author}{\bibfnamefont{Y.}~\bibnamefont{Nagamine}}, \bibnamefont{and}
  \bibinfo{author}{\bibfnamefont{D.}~\bibnamefont{Djayaprawira}},
  \bibinfo{journal}{J. Appl. Phys.} \textbf{\bibinfo{volume}{101}},
  \bibinfo{pages}{013907} (\bibinfo{year}{2007}).

\bibitem[{\citenamefont{Sun}(2015)}]{2015095}
\bibinfo{author}{\bibfnamefont{J.~Z.} \bibnamefont{Sun}},
  \bibinfo{journal}{Phys. Rev. B} \textbf{\bibinfo{volume}{91}},
  \bibinfo{pages}{174429} (\bibinfo{year}{2015}).

\bibitem[{\citenamefont{Shaw et~al.}(2015)\citenamefont{Shaw, Numbach, Weiler,
  Silva, martin Schoen, Sun, and Worledge}}]{2019010}
\bibinfo{author}{\bibfnamefont{J.~M.} \bibnamefont{Shaw}},
  \bibinfo{author}{\bibfnamefont{H.~T.} \bibnamefont{Numbach}},
  \bibinfo{author}{\bibfnamefont{M.}~\bibnamefont{Weiler}},
  \bibinfo{author}{\bibfnamefont{T.~J.} \bibnamefont{Silva}},
  \bibinfo{author}{\bibnamefont{martin Schoen}},
  \bibinfo{author}{\bibfnamefont{J.~Z.} \bibnamefont{Sun}}, \bibnamefont{and}
  \bibinfo{author}{\bibfnamefont{D.~C.} \bibnamefont{Worledge}},
  \bibinfo{journal}{IEEE Magn. Lett.} \textbf{\bibinfo{volume}{6}},
  \bibinfo{pages}{3500404} (\bibinfo{year}{2015}).

\bibitem[{\citenamefont{Sun}(2017)}]{Sun2017}
\bibinfo{author}{\bibfnamefont{J.~Z.} \bibnamefont{Sun}},
  \bibinfo{journal}{Phys. Rev. B} \textbf{\bibinfo{volume}{96}},
  \bibinfo{pages}{1} (\bibinfo{year}{2017}).

\bibitem[{\citenamefont{Sun et~al.}(2013)\citenamefont{Sun, Brown, Chen,
  Delenia, Gaidis, Harms, Hu, Jiang, Kilaru, Kula et~al.}}]{Sun2013}
\bibinfo{author}{\bibfnamefont{J.~Z.} \bibnamefont{Sun}},
  \bibinfo{author}{\bibfnamefont{S.~L.} \bibnamefont{Brown}},
  \bibinfo{author}{\bibfnamefont{W.}~\bibnamefont{Chen}},
  \bibinfo{author}{\bibfnamefont{E.~A.} \bibnamefont{Delenia}},
  \bibinfo{author}{\bibfnamefont{M.~C.} \bibnamefont{Gaidis}},
  \bibinfo{author}{\bibfnamefont{J.}~\bibnamefont{Harms}},
  \bibinfo{author}{\bibfnamefont{G.}~\bibnamefont{Hu}},
  \bibinfo{author}{\bibfnamefont{X.}~\bibnamefont{Jiang}},
  \bibinfo{author}{\bibfnamefont{R.}~\bibnamefont{Kilaru}},
  \bibinfo{author}{\bibfnamefont{W.}~\bibnamefont{Kula}}, \bibnamefont{et~al.},
  \bibinfo{journal}{Phys. Rev. B} \textbf{\bibinfo{volume}{88}},
  \bibinfo{pages}{1} (\bibinfo{year}{2013}).

\bibitem[{\citenamefont{Safranski et~al.}(2016)\citenamefont{Safranski, Chen,
  Krivorotov, and Sun}}]{Safranski2016}
\bibinfo{author}{\bibfnamefont{C.~J.} \bibnamefont{Safranski}},
  \bibinfo{author}{\bibfnamefont{Y.-J.} \bibnamefont{Chen}},
  \bibinfo{author}{\bibfnamefont{I.~N.} \bibnamefont{Krivorotov}},
  \bibnamefont{and} \bibinfo{author}{\bibfnamefont{J.~Z.} \bibnamefont{Sun}},
  \bibinfo{journal}{Appl. Phys. Lett.} \textbf{\bibinfo{volume}{109}},
  \bibinfo{pages}{132408} (\bibinfo{year}{2016}).

\bibitem[{\citenamefont{Kiselev et~al.}(2003)\citenamefont{Kiselev, Sankey,
  Krivorotov, Emley, Schoelkopf, Buhrman, and Ralph}}]{Kiselev2003}
\bibinfo{author}{\bibfnamefont{S.~I.} \bibnamefont{Kiselev}},
  \bibinfo{author}{\bibfnamefont{J.~C.} \bibnamefont{Sankey}},
  \bibinfo{author}{\bibfnamefont{I.~N.} \bibnamefont{Krivorotov}},
  \bibinfo{author}{\bibfnamefont{N.~C.} \bibnamefont{Emley}},
  \bibinfo{author}{\bibfnamefont{R.~J.} \bibnamefont{Schoelkopf}},
  \bibinfo{author}{\bibfnamefont{R.~a.} \bibnamefont{Buhrman}},
  \bibnamefont{and} \bibinfo{author}{\bibfnamefont{D.~C.} \bibnamefont{Ralph}},
  \bibinfo{journal}{Nature} \textbf{\bibinfo{volume}{425}},
  \bibinfo{pages}{380} (\bibinfo{year}{2003}).

\bibitem[{\citenamefont{Demidov et~al.}(2012)\citenamefont{Demidov, Urazhdin,
  Ulrichs, Tiberkevich, Slavin, Baither, Schmitz, and
  Demokritov}}]{Demidov2012}
\bibinfo{author}{\bibfnamefont{V.~E.} \bibnamefont{Demidov}},
  \bibinfo{author}{\bibfnamefont{S.}~\bibnamefont{Urazhdin}},
  \bibinfo{author}{\bibfnamefont{H.}~\bibnamefont{Ulrichs}},
  \bibinfo{author}{\bibfnamefont{V.}~\bibnamefont{Tiberkevich}},
  \bibinfo{author}{\bibfnamefont{A.}~\bibnamefont{Slavin}},
  \bibinfo{author}{\bibfnamefont{D.}~\bibnamefont{Baither}},
  \bibinfo{author}{\bibfnamefont{G.}~\bibnamefont{Schmitz}}, \bibnamefont{and}
  \bibinfo{author}{\bibfnamefont{S.~O.} \bibnamefont{Demokritov}},
  \bibinfo{journal}{Nat. Mater} \textbf{\bibinfo{volume}{11}},
  \bibinfo{pages}{1028} (\bibinfo{year}{2012}).

\bibitem[{\citenamefont{Safranski et~al.}(2017)\citenamefont{Safranski,
  Barsukov, Lee, Schneider, Jara, Smith, Chang, Lenz, Lindner, Tserkovnyak
  et~al.}}]{Safranski2017}
\bibinfo{author}{\bibfnamefont{C.}~\bibnamefont{Safranski}},
  \bibinfo{author}{\bibfnamefont{I.}~\bibnamefont{Barsukov}},
  \bibinfo{author}{\bibfnamefont{H.~K.} \bibnamefont{Lee}},
  \bibinfo{author}{\bibfnamefont{T.}~\bibnamefont{Schneider}},
  \bibinfo{author}{\bibfnamefont{A.~A.} \bibnamefont{Jara}},
  \bibinfo{author}{\bibfnamefont{A.}~\bibnamefont{Smith}},
  \bibinfo{author}{\bibfnamefont{H.}~\bibnamefont{Chang}},
  \bibinfo{author}{\bibfnamefont{K.}~\bibnamefont{Lenz}},
  \bibinfo{author}{\bibfnamefont{J.}~\bibnamefont{Lindner}},
  \bibinfo{author}{\bibfnamefont{Y.}~\bibnamefont{Tserkovnyak}},
  \bibnamefont{et~al.}, \bibinfo{journal}{Nat. Commun.}
  \textbf{\bibinfo{volume}{8}}, \bibinfo{pages}{117} (\bibinfo{year}{2017}).

\bibitem[{\citenamefont{Slavin and Tiberkevich}(2009)}]{Slavin2009}
\bibinfo{author}{\bibfnamefont{A.}~\bibnamefont{Slavin}} \bibnamefont{and}
  \bibinfo{author}{\bibfnamefont{V.}~\bibnamefont{Tiberkevich}},
  \bibinfo{journal}{IEEE Trans. Magn.} \textbf{\bibinfo{volume}{45}},
  \bibinfo{pages}{1875} (\bibinfo{year}{2009}).

\bibitem[{\citenamefont{Naletov et~al.}(2011)\citenamefont{Naletov, {De
  Loubens}, Albuquerque, Borlenghi, Cros, Faini, Grollier, Hurdequint,
  Locatelli, Pigeau et~al.}}]{Naletov2011}
\bibinfo{author}{\bibfnamefont{V.~V.} \bibnamefont{Naletov}},
  \bibinfo{author}{\bibfnamefont{G.}~\bibnamefont{{De Loubens}}},
  \bibinfo{author}{\bibfnamefont{G.}~\bibnamefont{Albuquerque}},
  \bibinfo{author}{\bibfnamefont{S.}~\bibnamefont{Borlenghi}},
  \bibinfo{author}{\bibfnamefont{V.}~\bibnamefont{Cros}},
  \bibinfo{author}{\bibfnamefont{G.}~\bibnamefont{Faini}},
  \bibinfo{author}{\bibfnamefont{J.}~\bibnamefont{Grollier}},
  \bibinfo{author}{\bibfnamefont{H.}~\bibnamefont{Hurdequint}},
  \bibinfo{author}{\bibfnamefont{N.}~\bibnamefont{Locatelli}},
  \bibinfo{author}{\bibfnamefont{B.}~\bibnamefont{Pigeau}},
  \bibnamefont{et~al.}, \bibinfo{journal}{Phys. Rev. B}
  \textbf{\bibinfo{volume}{84}}, \bibinfo{pages}{1} (\bibinfo{year}{2011}).

\bibitem[{\citenamefont{Gon{\c{c}}alves
  et~al.}(2013)\citenamefont{Gon{\c{c}}alves, Barsukov, Chen, Yang, Katine, and
  Krivorotov}}]{Goncalves2013}
\bibinfo{author}{\bibfnamefont{A.~M.} \bibnamefont{Gon{\c{c}}alves}},
  \bibinfo{author}{\bibfnamefont{I.}~\bibnamefont{Barsukov}},
  \bibinfo{author}{\bibfnamefont{Y.~J.} \bibnamefont{Chen}},
  \bibinfo{author}{\bibfnamefont{L.}~\bibnamefont{Yang}},
  \bibinfo{author}{\bibfnamefont{J.~A.} \bibnamefont{Katine}},
  \bibnamefont{and} \bibinfo{author}{\bibfnamefont{I.~N.}
  \bibnamefont{Krivorotov}}, \bibinfo{journal}{Appl. Phys. Lett.}
  \textbf{\bibinfo{volume}{103}} (\bibinfo{year}{2013}).

\bibitem[{\citenamefont{Tulapurkar et~al.}(2005)\citenamefont{Tulapurkar,
  Suzuki, Fukushima, Kubota, Maehara, Tsunekawa, Djayaprawira, Watanabe, and
  Yuasa}}]{Tulapurkar2005}
\bibinfo{author}{\bibfnamefont{A.~A.} \bibnamefont{Tulapurkar}},
  \bibinfo{author}{\bibfnamefont{Y.}~\bibnamefont{Suzuki}},
  \bibinfo{author}{\bibfnamefont{A.}~\bibnamefont{Fukushima}},
  \bibinfo{author}{\bibfnamefont{H.}~\bibnamefont{Kubota}},
  \bibinfo{author}{\bibfnamefont{H.}~\bibnamefont{Maehara}},
  \bibinfo{author}{\bibfnamefont{K.}~\bibnamefont{Tsunekawa}},
  \bibinfo{author}{\bibfnamefont{D.~D.} \bibnamefont{Djayaprawira}},
  \bibinfo{author}{\bibfnamefont{N.}~\bibnamefont{Watanabe}}, \bibnamefont{and}
  \bibinfo{author}{\bibfnamefont{S.}~\bibnamefont{Yuasa}},
  \bibinfo{journal}{Nature} \textbf{\bibinfo{volume}{438}},
  \bibinfo{pages}{339} (\bibinfo{year}{2005}).

\bibitem[{\citenamefont{Sankey et~al.}(2006)\citenamefont{Sankey, Braganca,
  Garcia, Krivorotov, Buhrman, and Ralph}}]{Sankey2006}
\bibinfo{author}{\bibfnamefont{J.~C.} \bibnamefont{Sankey}},
  \bibinfo{author}{\bibfnamefont{P.~M.} \bibnamefont{Braganca}},
  \bibinfo{author}{\bibfnamefont{A.~G.~F.} \bibnamefont{Garcia}},
  \bibinfo{author}{\bibfnamefont{I.~N.} \bibnamefont{Krivorotov}},
  \bibinfo{author}{\bibfnamefont{R.~A.} \bibnamefont{Buhrman}},
  \bibnamefont{and} \bibinfo{author}{\bibfnamefont{D.~C.} \bibnamefont{Ralph}},
  \bibinfo{journal}{Phys. Rev. Lett.} \textbf{\bibinfo{volume}{96}},
  \bibinfo{pages}{50} (\bibinfo{year}{2006}).

\bibitem[{\citenamefont{Petit et~al.}(2007)\citenamefont{Petit, Baraduc,
  Thirion, Ebels, Liu, Li, Wang, and Dieny}}]{Petit2007}
\bibinfo{author}{\bibfnamefont{S.}~\bibnamefont{Petit}},
  \bibinfo{author}{\bibfnamefont{C.}~\bibnamefont{Baraduc}},
  \bibinfo{author}{\bibfnamefont{C.}~\bibnamefont{Thirion}},
  \bibinfo{author}{\bibfnamefont{U.}~\bibnamefont{Ebels}},
  \bibinfo{author}{\bibfnamefont{Y.}~\bibnamefont{Liu}},
  \bibinfo{author}{\bibfnamefont{M.}~\bibnamefont{Li}},
  \bibinfo{author}{\bibfnamefont{P.}~\bibnamefont{Wang}}, \bibnamefont{and}
  \bibinfo{author}{\bibfnamefont{B.}~\bibnamefont{Dieny}},
  \bibinfo{journal}{Phys. Rev. Lett.} \textbf{\bibinfo{volume}{98}},
  \bibinfo{pages}{3} (\bibinfo{year}{2007}).

\bibitem[{\citenamefont{Muduli et~al.}(2012)\citenamefont{Muduli, Heinonen, and
  {\AA}kerman}}]{Muduli2012}
\bibinfo{author}{\bibfnamefont{P.~K.} \bibnamefont{Muduli}},
  \bibinfo{author}{\bibfnamefont{O.~G.} \bibnamefont{Heinonen}},
  \bibnamefont{and}
  \bibinfo{author}{\bibfnamefont{J.}~\bibnamefont{{\AA}kerman}},
  \bibinfo{journal}{Phys. Rev. Lett.} \textbf{\bibinfo{volume}{108}},
  \bibinfo{pages}{207203} (\bibinfo{year}{2012}).

\bibitem[{\citenamefont{Strelkov et~al.}(2018)\citenamefont{Strelkov, Chavent,
  Timopheev, Sousa, Prejbeanu, Buda-Prejbeanu, and Dieny}}]{Strelkov2018}
\bibinfo{author}{\bibfnamefont{N.}~\bibnamefont{Strelkov}},
  \bibinfo{author}{\bibfnamefont{A.}~\bibnamefont{Chavent}},
  \bibinfo{author}{\bibfnamefont{A.}~\bibnamefont{Timopheev}},
  \bibinfo{author}{\bibfnamefont{R.~C.} \bibnamefont{Sousa}},
  \bibinfo{author}{\bibfnamefont{I.~L.} \bibnamefont{Prejbeanu}},
  \bibinfo{author}{\bibfnamefont{L.~D.} \bibnamefont{Buda-Prejbeanu}},
  \bibnamefont{and} \bibinfo{author}{\bibfnamefont{B.}~\bibnamefont{Dieny}},
  \bibinfo{journal}{Phys. Rev. B} \textbf{\bibinfo{volume}{98}},
  \bibinfo{pages}{214410} (\bibinfo{year}{2018}).

\bibitem[{\citenamefont{Ando et~al.}(2009)\citenamefont{Ando, Toda, Nozaki,
  Mizukami, Maruyama, Ohta, Tulapurkar, Shiraishi, Mizuguchi, Shinjo
  et~al.}}]{Ando2009}
\bibinfo{author}{\bibfnamefont{Y.}~\bibnamefont{Ando}},
  \bibinfo{author}{\bibfnamefont{N.}~\bibnamefont{Toda}},
  \bibinfo{author}{\bibfnamefont{T.}~\bibnamefont{Nozaki}},
  \bibinfo{author}{\bibfnamefont{S.}~\bibnamefont{Mizukami}},
  \bibinfo{author}{\bibfnamefont{T.}~\bibnamefont{Maruyama}},
  \bibinfo{author}{\bibfnamefont{K.}~\bibnamefont{Ohta}},
  \bibinfo{author}{\bibfnamefont{A.~A.} \bibnamefont{Tulapurkar}},
  \bibinfo{author}{\bibfnamefont{M.}~\bibnamefont{Shiraishi}},
  \bibinfo{author}{\bibfnamefont{M.}~\bibnamefont{Mizuguchi}},
  \bibinfo{author}{\bibfnamefont{T.}~\bibnamefont{Shinjo}},
  \bibnamefont{et~al.}, \bibinfo{journal}{Nat. Nanotechnol.}
  \textbf{\bibinfo{volume}{4}}, \bibinfo{pages}{158} (\bibinfo{year}{2009}).

\bibitem[{\citenamefont{Dohi et~al.}(2017)\citenamefont{Dohi, Kanai, Matsukura,
  and Ohno}}]{Dohi2017}
\bibinfo{author}{\bibfnamefont{T.}~\bibnamefont{Dohi}},
  \bibinfo{author}{\bibfnamefont{S.}~\bibnamefont{Kanai}},
  \bibinfo{author}{\bibfnamefont{F.}~\bibnamefont{Matsukura}},
  \bibnamefont{and} \bibinfo{author}{\bibfnamefont{H.}~\bibnamefont{Ohno}},
  \bibinfo{journal}{Appl. Phys. Lett.} \textbf{\bibinfo{volume}{111}},
  \bibinfo{pages}{072403} (\bibinfo{year}{2017}).

\bibitem[{\citenamefont{Auerbach et~al.}(2018)\citenamefont{Auerbach, Leder,
  Gider, and Arthaber}}]{Auerbach2018}
\bibinfo{author}{\bibfnamefont{E.}~\bibnamefont{Auerbach}},
  \bibinfo{author}{\bibfnamefont{N.}~\bibnamefont{Leder}},
  \bibinfo{author}{\bibfnamefont{S.}~\bibnamefont{Gider}}, \bibnamefont{and}
  \bibinfo{author}{\bibfnamefont{H.}~\bibnamefont{Arthaber}},
  \bibinfo{journal}{IEEE Transactions on Magnetics}
  \textbf{\bibinfo{volume}{54}}, \bibinfo{pages}{1} (\bibinfo{year}{2018}).

\bibitem[{\citenamefont{Carcia et~al.}(1985)\citenamefont{Carcia, Meinhaldt,
  and Suna}}]{Carcia1985}
\bibinfo{author}{\bibfnamefont{P.~F.} \bibnamefont{Carcia}},
  \bibinfo{author}{\bibfnamefont{A.~D.} \bibnamefont{Meinhaldt}},
  \bibnamefont{and} \bibinfo{author}{\bibfnamefont{A.}~\bibnamefont{Suna}},
  \bibinfo{journal}{Appl. Phys. Lett.} \textbf{\bibinfo{volume}{47}},
  \bibinfo{pages}{178} (\bibinfo{year}{1985}).

\end{thebibliography}

\end{document}